\documentclass[a4paper]{article}
\usepackage{multirow}
\usepackage{makecell}
\usepackage{boldline}
\usepackage{stfloats}
\usepackage{array}
\usepackage{hanging}
\usepackage{amsmath}
\usepackage{flushend}
\newcolumntype{C}[1]{>{\centering\let\newline\\\arraybackslash\hspace{0pt}}m{#1}}

\usepackage[table]{xcolor}
\definecolor{maroon}{cmyk}{0.5,0.5,0.5,0.0}
\definecolor{Gray}{gray}{0.9}
\usepackage{INTERSPEECH2021}

\title{Do You Listen with One or Two Microphones? A Unified ASR Model for Single and Multi-Channel Audio}
\name{Gokce Keskin, Minhua Wu, Brian King, Harish Mallidi, Yang Gao, \\Jasha Droppo, Ariya Rastrow, Roland Maas}
\address{
	Amazon.com
	}
\email{\{gkeskin, wuminhua, bbking, mallidih, ygaa, drojasha, arastrow, rmaas\}@amazon.com}

\begin{document}
	
	\maketitle
	\begin{abstract}
		Automatic speech recognition (ASR) models are typically designed to operate on a single input data type, e.g. a single or multi-channel audio streamed from a device. This design decision assumes the \textit{primary} input data source does not change and if an additional (\textit{auxiliary}) data source is occasionally available, it cannot be used. 
		An ASR model that operates on both primary and auxiliary data can achieve better accuracy compared to a primary-only solution; and a model that can serve both \textit{primary-only} (PO) and \textit{primary-plus-auxiliary} (PPA) modes is highly desirable. In this work, we propose a unified ASR model that can serve both modes. We demonstrate its efficacy in a realistic scenario where a set of devices typically stream a single primary audio channel, and two additional auxiliary channels \textit{only when} upload bandwidth allows it. The architecture enables a unique methodology that uses both types of input audio during training time. Our proposed approach achieves up to 12.5\% relative word-error-rate reduction (WERR) compared to a PO baseline, and up to 16.0\% relative WERR in low-SNR conditions. The unique training methodology achieves up to 2.5\% relative WERR compared to a PPA baseline.
		
	\end{abstract}
	
	\section{Introduction}

	ASR models are typically designed to assume that all input sources to the model are always available for each sample. The inputs could be acoustic data from a single audio channel, from multiple channels, or acoustic data combined with context vector embeddings from prior utterances in a conversational setting \cite{rnnt, minhua2019, Kim2019}. This design choice prevents the ASR model to accept additional input sources that contain useful information but are only \textit{occasionally} available. For instance, consider a classroom scenario where there is a central listener device with a microphone array and an additional lapel microphone that is occasionally used by the teacher. The input to the ASR could be a single audio channel that comes from an on-device beamformer in the central listener. If the upload bandwidth allows it, additional raw microphone channels (auxiliary inputs) could also be streamed. If the lapel is used, another auxiliary source is available. One would need to build separate ASR models for these scenarios with their own datasets. 
	
	In this work, we propose a unified model that can serve all these predefined scenarios. The unified model has separate frontends for each scenario (Sec. \ref{sec:architecture}) and employs a unique training methodology that combines datasets with different number of sources (Sec. \ref{sec:training}). In the rest of this paper, we present the results of such a unified model in a far-field ASR scenario: a wide variety of devices stream single-channel (SC) audio, but a subset of them might conditionally stream additional raw audio from microphones to create a multi-channel (MC) input. The unified model, coupled with the proposed training methodology, leads to lower WER than building an MC-only model that can only be trained with MC audio (Sec. \ref{sec:results}).
	\begin{figure}[t!]
		\centering
		\includegraphics[width=0.85\linewidth]{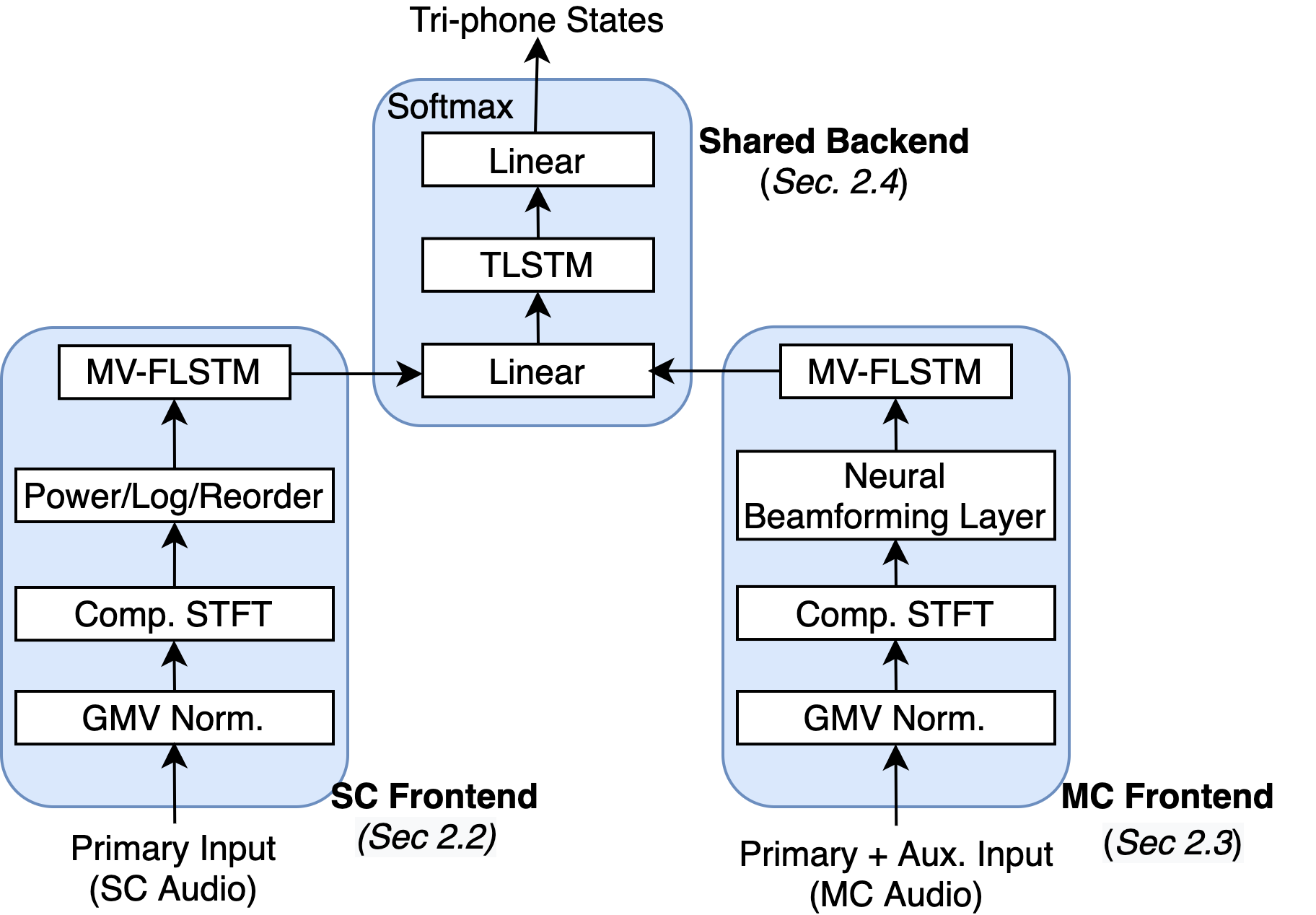}
		\caption{A Unified ASR model architecture. Separate frontends for each input type (primary and primary+auxiliary) share a backend, enabling a single model that serves both data types.}
		\label{fig:architecture}
	\end{figure}

	\section{Related Work}
	\label{sec:related-work}
	 Far-field ASR systems are designed to operate in more challenging acoustic conditions compared to a near-field system where the speaker is close to the microphone. Lower signal-to-noise ratio (SNR) in the received signal in the microphones reduces the word error rate (WER) of the following ASR system. Increased signal degradation with distance, room reverberation, noise, and background speech contribute to this reduction \cite{umbach2021}. 
	 
	 A complete distant speech recognition (DSR) system  typically consists of distinct components such as a voice activity detector (VAD), speaker localizer (SL), dereverberator, beamformer and acoustic model \cite{Omolog2001,Wolfel2009,KumataniAYMRST12,KinoshitaDGHHKL16,VirtanenBook2012}. Beamforming techniques take advantage of multiple microphones to enhance the audio signal, and is a key component to improve noise robustness of the DSR. Beamforming can be categorized into fixed beamforming or adaptive beamforming. In comparison to fixed beamforming, adaptive techniques have shown that noise robustness of ASR system can be improved with a dereverberation approach or high-order statistics. However, adaptive techniques rely on accurate VAD or SL, and therefore they can underperform in comparison to a fixed beam former; especially when these dependent components are not performing reliably. According to previous studies, individually optimizing various DSR components is sub-optimal \cite{McDonough2008, Seltzer2008}. 
	 
	 More recently, multi-channel deep neural network (MC-DNN) approaches have been applied to ASR by training a unified MC-DNN model where the MC processing modules are part of the DNN structure \cite{xiao2016deep,ochiai2017multichannel,sainath2017multichannel,minhua2019frequency}. Aside from unified MC-DNN approaches, a DNN is also employed to construct a clean speech signal. A mask-based method was proposed to estimate the statistics of the target clean speech via an LSTM \cite{heymann2018performance, higuchi2018frame}.  However, this method needs accumulated statistics from adequate amount of adaptation data to perrform well. Accumulating the statistics might cause additional latency and is less applicable to real-time applications.

\section{Model Architecture}
\label{sec:architecture}

\subsection{Overview}
\label{subsec:overview}
The unified architecture diagram is given in Fig. \ref{fig:architecture}. The model includes two separate frontends for SC and MC audio, with a shared backend. SC audio, or the \textit{primary channel}, is either obtained from the only existing microphone in the device or from an on-device beamformer that combines the outputs of multiple microphones. In our experiments, MC audio has three channels: the primary channel and two \textit{auxiliary channels} that are obtained from the raw audio outputs of two microphones.

The model is trained with a mix of SC and MC audio. MC samples propagate through the MC frontend (FE) and the shared backend, whereas SC samples propagate through the SC FE and the shared backend. During inference, audio is propagated through the corresponding FE for the incoming data type, followed by the shared backend (Sec. \ref{sec:training}).

\subsection{SC Frontend}
\label{subsec:sc-frontend}
In the SC FE, extracted audio features (Sec. \ref{subsubsec:fex}) are  processed by a multi-view frequency LSTM (MV-FLSTM) \cite{maarten_flstm} (Sec. \ref{subsubsec:sc-mv-flstm}), followed by a shared backend (Sec. \ref{subsec:backend}).

\subsubsection{Feature Extraction}
\label{subsubsec:fex}
Global mean and variance (GMV) values computed from the received channel are used to normalize the input. Complex STFT features with 256 frequency bins are extracted from the normalized  waveform with a window size of 25ms produced in 10ms steps. Three consecutive input frames are stacked into a single vector to reduce the number of time steps in the acoustic model, known as the Lower Frame Rate model \cite{lfr2016}. Log-power of the complex features is consumed by the MV-FLSTM (Sec. \ref{subsubsec:sc-mv-flstm}).
\begin{figure}[t!]
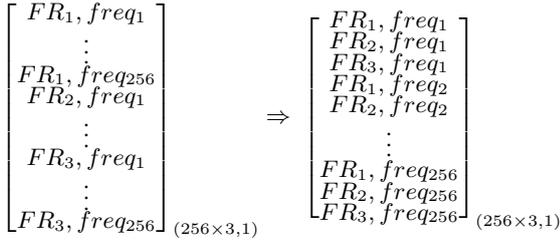

	\centering
	\renewcommand{\arraystretch}{0.2}
	${
		\begin{bmatrix}
			FR_1, freq_1 \\
			\vdots \\
			FR_1, freq_{256}\\
			FR_2, freq_1\\
			\vdots \\
			FR_3, freq_1\\
			\vdots \\
			FR_3, freq_{256}
		\end{bmatrix}_{(256\times3,1)}
	}$ 
	$\Rightarrow$ 
	\renewcommand{\arraystretch}{0.5}
	${
		\begin{bmatrix}
			FR_1, freq_1 \\
			FR_2, freq_1 \\
			FR_3, freq_1\\
			FR_1, freq_2 \\
			FR_2, freq_2\\
			\vdots \\
			FR_1, freq_{256}\\
			FR_2, freq_{256}\\
			FR_3, freq_{256}
		\end{bmatrix}_{(256\times3,1)}
	}$
	\caption{Input to the SC MV-FLSTM is reordered to put frequency bins closer. A similar procedure is done before the MC MV-FLSTM, incorporating the look directions in the reorder.}
	\label{fig:mvflstm2}
\end{figure}
\subsubsection{SC Frontend MV-FLSTM}
\label{subsubsec:sc-mv-flstm}
A Frequency LSTM (FLSTM) operates over the frequencies contained in an individual input frame with a sliding, overlapping window \cite{flstmli}. Multi-view FLSTM extends this concept by having several window sizes (\textit{views}) to span different frequency ranges at each step \cite{maarten_flstm}. Outputs of the views are concatenated to generate a feature vector that is later processed by a conventional time-domain LSTM (TLSTM). Concatenated feature vector can optionally be projected to a lower dimensional space to reduce computational requirements of the following TLSTM. Prior work has shown 3\%-7\% relative gains in WER by using the MV-FLSTM approach over a single-view FLSTM \cite{maarten_flstm}.

Since FLSTMs operate over the frequency dimension, the output of the feature extraction stage consisting of three stacked frames is reordered. This reordering ensures the same frequency bins are clustered together before being processed by the MV-FLSTM (Fig. \ref{fig:mvflstm2}). In this paper, four FLSTM views are used. Each view consists of a three-layer bidirectional FLSTM with 32 cells in each layer. FLSTMs operate on data that is contained in a single frame, hence bidirectionality does not look ahead in time and disrupt causality. Window sizes of the views are [24, 48, 96, 192] and hop size is half of the window size. 

\subsection {MC Frontend}
MC audio is normalized by GMV statistics for each channel. Neural Beamforming Layer (NBL) combines the audio channels and produces a number of look directions, which are processed by a separate MV-FLSTM and the shared backend. 
\begin{figure}[t!]
	\centering
	\includegraphics[width=0.35\linewidth]{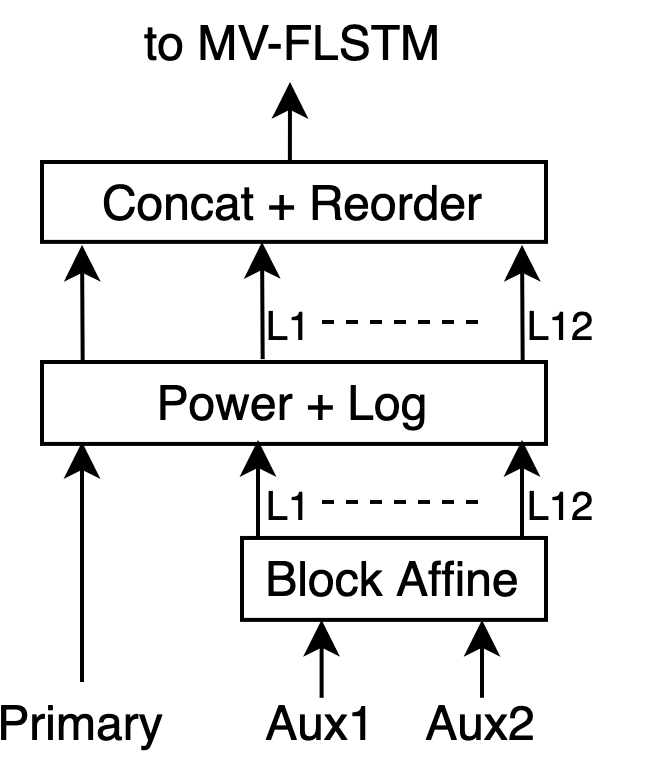}
	\caption{Neural Beamforming Layer.}
	\label{fig:beamformer}
\end{figure}
\subsubsection{Neural Beamforming Layer}
\label{subsubsec:beamformer}

NBL implementation is adopted from the Elastic Spatial Filter described in \cite{minhua2019}. The Discrete Fourier Transform of the normalized input signal can be described as below:
\begin{equation}
	\mathbf{X}\left(t, \omega_{k}\right)=\left[X_{1}\left(t, \omega_{k}\right), \cdots, X_{M}\left(t, \omega_{k}\right)\right]^{T}
\end{equation}
Using this notation, we can express the complex weight vector for source position $\mathbf{p}$ as follows:
\begin{equation}
	\mathbf{w}^H\left(t, \omega_{k}, \mathbf{p}\right)=\left[w_{1}\left(t, \omega_{k}, \mathbf{p}\right), \cdots, w_{M}\left(t, \omega_{k}, \mathbf{p}\right)\right]
\end{equation}
Thus, the block affine transform (BAT) can be expressed as:
\begin{equation}\footnotesize
	\left[\begin{array}{c}{Y_{1}\left(\omega_{1}\right)} \\ {\dots} \\ {Y_{D}\left(\omega_{1}\right)} \\ {\dots} \\ {Y_{1}\left(\omega_{K}\right)} \\ {\dots} \\ {Y_{D}\left(\omega_{K}\right)}\end{array}\right] = \left[\begin{array}{c}{\mathbf{w}_{\mathrm{SD}}^{H}\left(\omega_{1}, \mathbf{p}_{1}\right) \mathbf{X}\left(\omega_{1}\right)+\mathbf{b}_{1}} \\ {\dots} \\ {\mathbf{w}_{\mathrm{SD}}^{H}\left(\omega_{1}, \mathbf{p}_{D}\right) \mathbf{X}\left(\omega_{1}\right)+\mathbf{b}_{D}} \\ {\dots} \\ {\mathbf{w}_{\mathrm{SD}}^{H}\left(\omega_{K}, \mathbf{p}_{1}\right) \mathbf{X}\left(\omega_{K}\right)+\mathbf{b}_{D (K-1)+1}} \\ {\dots} \\ {\mathbf{w}_{\mathrm{SD}}^{H}\left(\omega_{K}, \mathbf{p}_{D}\right) \mathbf{X}\left(\omega_{K}\right)+\mathbf{b}_{DK}}\end{array}\right]
	\normalsize
\end{equation} where  $\mathbf{b}$ is bias term, $D$ is the number of look directions and $K$ is the number of frequency bins.

Complex STFT features from the two microphones are processed by the trainable BAT, generating 12 look directions (Fig. \ref{fig:beamformer}). BAT is initialized with super directive beamformer weights. Log-power features of the look directions and the primary channel are concatenated and further processed by the MC FE's MV-FLSTM. Early experiments showed that using all three channels in the MC FE had 4-7\% relative WERR compared to using only the two auxiliary (raw) channels, hence the primary input is concetanated to the BAT output.

\subsubsection{MC Frontend MV-FLSTM}
\label{subsubsec:mc-mv-flstm}
The input to the MC FE's MV-FLSTM is 13 times the size of the input to the SC FE's MV-FLSTM. This is due to the concatenation of the primary channel and the look directions from the NBL. In order to ensure the same range of frequencies are spanned, each view in MC FE's MV-FLSTM has 13 times the window and hop size compared to the SC FE implementation. The number of views, layers and units-per-layer are the same for both frontends. Similar to the SC FE implementation, the input to the MV-FLSTM is reordered and processed frequency bins from different look directions (in addition to the stacked frames) are arranged to be next to each other (Fig. \ref{fig:mvflstm2}).

\subsection{Shared Backend}
\label{subsec:backend}
The shared backend has a projection (linear) layer to reduce the input dimensionality to the five-layer, unidirectional TLSTM with 768 cells per layer. Unidirectionality is required to preserve causality for a streamable ASR model. TLSTM output is connected to a classification layer with softmax outputs to generate tied tri-phone states used in a hybrid ASR model \cite{denglibook}. Unified model has a total of 28M parameters.
\begin{figure}[t!]
	\centering
	\includegraphics[width=0.7\linewidth]{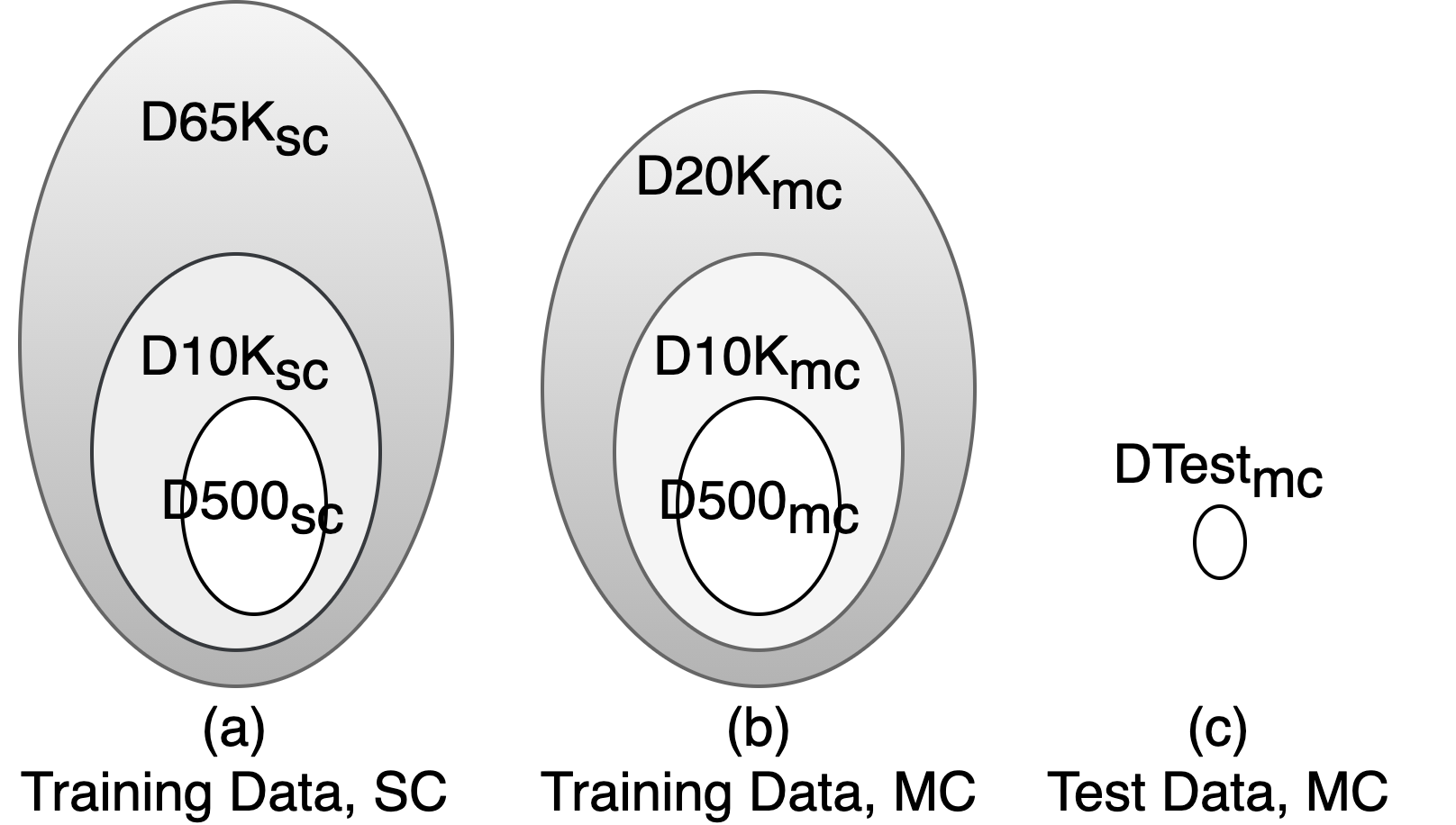}
	\caption{Datasets in this work: (a) SC training, with [500, 10K, 65K] hours, (b) MC training, with [500, 10K, 20K] hours, (c) Test, 45 hours. SC data has one primary channel, MC data has one primary and two auxiliary channels. Large datasets include smaller ones. MC and SC datasets don't overlap.}
	\label{fig:datasets}
\end{figure}

\section{Datasets}
\label{sec:datasets}
Datasets used in this work are shown in Fig. \ref{fig:datasets}. SC datasets D500\textsubscript{sc}, D10K\textsubscript{sc} and D65K\textsubscript{sc} are human transcribed with [500, 10K, 65K] hours of audio respectively and they contain only the primary channel. MC datasets D500\textsubscript{mc}, D10K\textsubscript{mc} and D20K\textsubscript{mc} have [500, 10K, 20K] hours of audio respectively and contain three channels: primary channel and two auxiliary channels obtained from raw audio of two microphones. MC datasets are transcribed by an accurate SC model that is free from production constraints such as latency, causality and size. 

DTest\textsubscript{mc} is a human-transcribed MC test set with 45 hours of audio, containing three channels (one primary and two auxiliary). For evaluation of SC models, only the primary channel in this set is used. The test set contains a mix of single and multi-speaker utterances (i.e., background speech). SNR value for each utterance is also available in the test set to evaluate performance across different noise conditions. All audio data is de-identified for privacy reasons.

%

\section{Training Methodology}
\label{sec:training}
During training, each batch of data contains a mix of MC and SC audio. Since the primary channel is present in the MC dataset, the auxiliary channels can be removed and the resulting SC data is added to expand the SC dataset. Gradient updates obtained from the expanded SC dataset are used to train the SC FE and the shared backend. Gradients generated from the MC data are used to update the MC FE and the shared backend. Notably, the unified ASR architecture described in Sec. \ref{sec:architecture} allows the shared backend to learn from both SC and MC data.

There are other alternatives to incorporate SC data to MC model training. One can pad the missing two channels in the SC data with zeros to expand the MC dataset and train an MC model with this expanded dataset. However, experiments show this is inferior to unified model training with two separate frontends (Sec. \ref{subsec:unified-vs-zeropad}). Another alternative is to first train an SC-only model to obtain the backend, freeze its weights, append an MC FE and then train the MC FE with MC-only data. Freezing the weights is required to enable a single model for SC and MC audio. This approach is undesirable since it requires a two-step methodology and the backend will not be updated for MC data. In practice, training in this manner led to non-convergence of the MC FE in large-scale datasets.

During inference of the unified ASR model, either the SC or the MC FE is used based on the incoming data type. 
\begin{table}[t!]
	\caption{Normalized WER Comparison of Unified Model with Two Frontends to Zero-Padding Missing Channels.}
	\label{tab:results0}
	\begin{tabular}{|C{0.9cm} | C{1.1cm} | C{2.5cm}| C{2.0cm}|}
		\hline
		Experiment & Training Data & Model / Inference Path & nWER (Single/Multi-Speaker)\\ \Xhline{4\arrayrulewidth}
		E1  & D500\textsubscript{sc} +D500\textsubscript{mc} & MC, Zero-pad missing channels & 100.0/100.0 \\ \hline
		E2 & D500\textsubscript{sc} +D500\textsubscript{mc} &  Unified (MC FE + Shared Backend) & \textbf{96.0}/\textbf{96.9}  \\ \hline
	\end{tabular}
	
\end{table}

\section{Experimental Results}
\label{sec:results}
Models are trained with cross-entropy (CE) loss, followed by CTC loss \cite{ctc} for the same number of training epochs as the corresponding baseline. WER results are obtained from the test set DTest\textsubscript{mc }(Sec. \ref{sec:datasets}) and normalized to the baseline. Absolute WER values are below 10\% for medium and large-scale datasets. Only the primary channel is used for SC-only models during training and test.

\subsection{Comparison of Proposed Approach to Zero Padding}
\label{subsec:unified-vs-zeropad}
Table \ref{tab:results0} shows the Normalized WER (nWER) of the zero-padding baseline (E1) to the proposed unified model architecture (E2). E1 is a combined MC/SC model that consists of the same MC FE and the shared backend as E2 (Sec. \ref{sec:architecture}), but does not include an SC FE. E1 always expects three input channels (one primary and two auxiliary); if only SC data is available during training or inference, the missing channels are filled with zeros. E2 is trained with two separate frontends (Sec. \ref{sec:training}), with each input type going through its respective frontend. 

During test time, all three channels in DTest\textsubscript{mc} are used as input to E1 and E2. WER values of E1 for single and multi-speaker cases are arbitrarily set to 100.0, and E2's WER values are normalized to E1. The unified model architecture achieves 3.1\%-4.0\% relative WERR compared to the baseline, demonstrating its advantage over the alternative method.

\begin{table*}[t!]
	\centering
	\caption{Normalized WER Comparison of MC and SC models.}
	\label{tab:results1}
	\begin{tabular}{|C{1.5cm} | C{2.8cm} | C{5.5cm}| C{3.5cm}|}
		\hline
		Experiment & Training Data & Model / Inference Path & { nWER \newline (Single/Multi-Speaker)} \\ \Xhline{4\arrayrulewidth}
		E3  & D10K\textsubscript{sc} + D10K\textsubscript{mc} & Standalone SC & 100.0/100.0 \\ \hline
		E4 & D10K\textsubscript{mc} &  Standalone MC & 89.7/94.8  \\ \hline
		E5  & D10K\textsubscript{sc} + D10K\textsubscript{mc}  & Unified (SC FE + Shared Backend) & 101.5/99.3 \\ \hline
		E6 & D10K\textsubscript{sc} + D10K\textsubscript{mc}  & Unified (MC FE + Shared Backend) & \textbf{87.5}/\textbf{92.3} \\ \hline \hline
		E7 & D65K\textsubscript{sc} + D20K\textsubscript{mc}  & Unified (SC FE + Shared Backend) & 85.9/88.7 \\ \hline
		E8 & D65K\textsubscript{sc} + D20K\textsubscript{mc}  & Unified (MC FE + Shared Backend) & \textbf{82.6}/\textbf{85.7} \\ \hline
	\end{tabular}

\end{table*}

\subsection{Impact of Including Multi-Channel Audio}
\label{subsec:results-mc}
Table \ref{tab:results1} shows the nWER comparison of the SC and MC models in medium and large-scale datasets. E3 has the same SC FE and the shared backend architecture in Fig. \ref{fig:architecture}, but does not contain the MC FE. Datasets D10K\textsubscript{sc} and D10K\textsubscript{mc} are combined for training (Sec. \ref{sec:datasets}); however E3 only uses the primary channel in D10K\textsubscript{mc} since it is an SC-only model. WER values of E3 for single and multi-speaker cases are arbitrarily set to 100.0, and following experiments' WER values are normalized to E3.

E4 is an MC model that combines the MC FE and the shared backend, and no SC FE. It is trained with D10K\textsubscript{mc} and uses all three channels. E3 and E4 are trained separately, with no weight sharing. E4 obtains a 10.3\%  relative WER reduction compared to the SC model (E3) in the single-speaker test set, even though E4 is trained with half the data of E3. This clearly demonstrates the improvements that can be obtained from using multiple audio channels. Multi-speaker test set has 5.2\% reduction.

\subsection{Impact of Unified Model Training}
\label{sec:umt}
Experiment E5 shows the results for the SC path (SC FE + shared backend) in the unified model and E6 shows the MC path results for the same model. E5 and E6 are trained together (Sec. \ref{sec:training}). Compared to E4, E6 achieves a further 2.2\%/2.5\% WERR in single/multi-speaker conditions. This demonstrates an advantage of the proposed architecture. 

E5 has a 0.7\% improvement in multi-speaker over E3, but a 1.5\% degradation in single-speaker conditions is observed. It is not clear if this degradation is an artifact of the unified model methodology or due to the inherent variability in training dynamics. Multiple runs with different initialization seeds might be warranted for further study.

\subsection{Impact of Large-Scale Data}
\label{subsec:results-lsd}
Table \ref{tab:results1} also shows the nWER comparison of the unified model with SC (E7) and MC (E8) paths in a large-scale data setting. E7 obtains 14.1\%/11.3\% reduction in relative WER compared to the SC baseline model E3. E8 has an additional 3.3\%/3.0\% relative gain over E7. SC data is more widely available due to the prevalence of SC ASR models, and our dataset distribution reflects this fact. Additional experiments with a different data distribution (e.g., oversampling the MC data to reach a 50/50 distribution) could determine if MC results can be further improved.

%

\subsection{Impact of Utterance SNR}
\label{subsec:snr}
\begin{figure}[t!]
	\centering
	\includegraphics[width=0.77\linewidth]{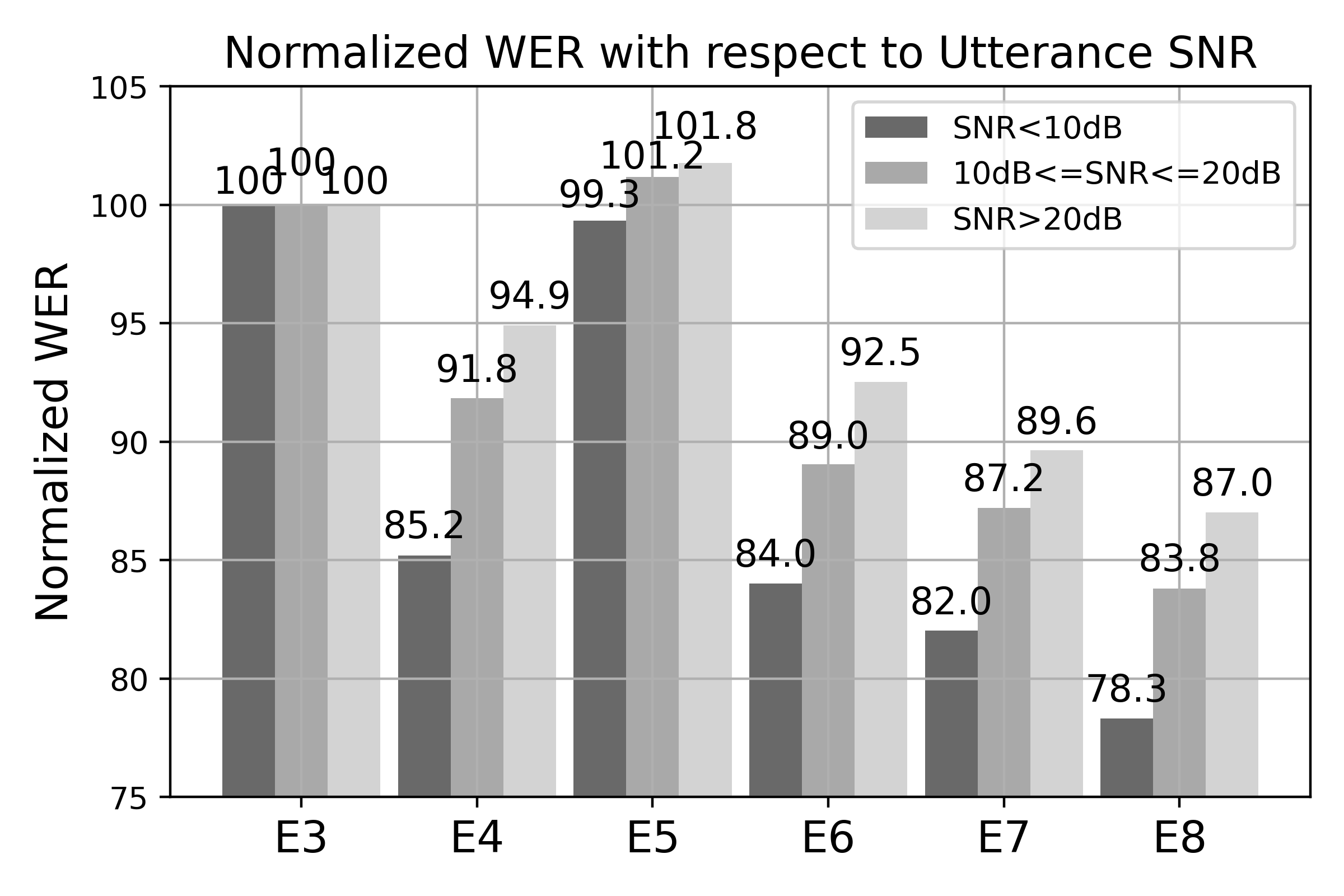}
	\caption{Relative WERR is more pronounced in low-SNR conditions when an MC model is used (e.g., E4 vs E3). Additional data helps more in low-SNR conditions (E7 vs. E5).}
	\label{fig:snr}
\end{figure}

Fig. \ref{fig:snr} shows the nWER for the models with respect to SNR of utterances. The test set is split into three bins according to individual utterances' SNR, and WER is computed for each bin. WER for each bin is normalized to the corresponding SNR bin of E3, whose WER is arbitrarily set to 100.0.

Comparing the MC model E4 to the SC model E3, the advantage of the MC FE is more pronounced in low-SNR ($<10dB$) conditions. A significant 14.8\% relative WERR is observed in this regime. This is perhaps not surprising since additional information available in the auxiliary channels is even more valuable in these noisy utterances.  Conversely, the advantage of incorporating SC data in E6 is more evident in medium and high-SNR conditions, with 2.8\% and 2.4\% relative WERR compared to E4. In low-SNR conditions, a smaller 1.2\% improvement is observed. E6 achieves 16.0\% relative WERR in low-SNR conditions compared to the SC baseline E3.

Adding large-scale data also helps significantly in low-SNR conditions. Comparing SC-models E5 and E7, a 17.3\% relative WERR is observed. In medium-SNR ($[10dB, 20dB]$) and high-SNR ($>20dB$) conditions, still significant reductions of 14.0\% and 12.2\% are seen. The combination of all three techniques (unified model, MC FE and additional data) leads to an impressive 21.7\% relative WERR in low-SNR conditions (E3 vs. E8), with 16.2\% and 13.0\% reductions in medium and high-SNR utterances.

\section{Conclusion}
We propose a unified MC/SC model that can be trained with both types of input, allowing a single model to support a variety of  scenarios. Proposed approach achieves up to 2.5\% relative WERR compared to the MC baseline and up to 16.0\% relative WERR compared to the SC baseline in low-SNR conditions.

\bibliographystyle{IEEEtran}
\bibliography{mybib}



\end{document}